\pdfoutput=1 
\documentclass[prd,twocolumn,nofootinbib,preprintnumbers,superscriptaddress,showpacs]{revtex4-1}

\usepackage{graphicx}
\usepackage{bm}
\usepackage{amsmath}
\usepackage{lineno}
\usepackage{amssymb}
\usepackage[utf8]{inputenc}
\usepackage[T1]{fontenc}
\usepackage{lipsum}
\usepackage{stmaryrd}
\usepackage{xcolor}
\usepackage[colorlinks,pdfstartview=FitH]{hyperref}
\hypersetup{linkcolor=blue,citecolor=blue,filecolor=black,urlcolor=blue}

\IfFileExists{dsfont.sty}
	{\usepackage{dsfont}
         \let\mathbb=\mathds
         \newcommand{\id}{\mathds{1}}}
	{\typeout{Package dsfont.sty was not found, using alternative macros.}
         \let\mathds=\mathbb
         \newcommand{\id}{\mbox{1 \kern-.59em {\rm l}}}}

\renewcommand\a{\alpha}
\renewcommand\b{\beta}
\renewcommand\d{\delta}
\renewcommand\k{\kappa}

\newcommand\g{\gamma}

\newcommand\m{\mu}

\newcommand\p{\pi}

\newcommand\s{\sigma}

\newcommand\D{\Delta}

\newcommand{\vare}{\mathrm{e}}

\renewcommand{\vec}{\boldsymbol}

\renewcommand{\part}{{\rm part}}

\newcommand{\be}{\begin{equation}}
\newcommand{\ee}{\end{equation}}
\newcommand{\bes}{\begin{subequations}}
\newcommand{\ees}{\end{subequations}}
\newcommand{\bea}{\begin{eqnarray}}
\newcommand{\eea}{\end{eqnarray}}

\newcommand{\na}{\nabla}

\def\nbox#1#2{\vcenter{\hrule \hbox{\vrule height#2in
\kern#1in \vrule} \hrule}}
\def\sq{\,\raise.5pt\hbox{$\nbox{.10}{.10}$}\,}
\def\sqb{\,\raise.5pt\hbox{$\overline{\nbox{.09}{.09}}$}\,}

\begin{document}
\author{Michael Lublinsky}
\email{lublinm@bgu.ac.il}
\affiliation{Department of Physics, Ben-Gurion University of the Negev,  Beer-Sheva 84105, Israel}
\author{Jared Reiten}
\email{jdreiten@physics.ucla.edu}
\affiliation{Department of Physics and Astronomy, University of California, Los Angeles,CA 90095, U.S.A.}
\affiliation{Mani L. Bhaumik Institute for Theoretical Physics,  University of California,  Los Angeles,  CA 90095, U.S.A.}
\author{Andrey V. Sadofyev}
\email{andrey.sadofyev@usc.es}
\affiliation{Instituto Galego de F{\'{i}}sica de Altas Enerx{\'{i}}as (IGFAE),  Universidade de Santiago de Compostela, Santiago de Compostela 15782,  Spain}
\affiliation{Institute for Theoretical and Experimental Physics (ITEP),  
Moscow 117218, Russia}

\title{Magnetic monopole in a chiral plasma: chiral dyon}

\begin{abstract}

The placement of a magnetic monopole into an electrically-neutral chiral plasma with a non-zero axial density results in an electric polarization of the matter.  The electric current produced by the chiral magnetic effect is balanced by charge diffusion and Ohmic dissipation,  which generates a non-trivial charge distribution.  In turn,  the latter induces a separation of chiralities along the magnetic field of the monopole due to the chiral separation effect.
We find the stationary states of such a system, with vanishing total electric current and stationary axial current balanced by the chiral anomaly.  In this solution,  the monopole becomes ``dressed'' with an electric charge that is proportional to the averaged chiral density of the matter --- forming a chiral dyon. The interplay between the chiral effects on the one hand, and presence of magnetic field of the monopole on the other, may affect the evolution of the monopole density in the early Universe,  contribute to the process of baryogenesis,  and can also be instrumental for detection of relic monopoles using chiral materials.

\end{abstract}

\maketitle

\section{Introduction}

Macroscopic manifestations of the axial anomaly have attracted significant attention in the literature, for a review see \cite{Kharzeev:2015znc, Huang:2015oca}. The corresponding transport phenomena, known as chiral effects, may take place in a variety of systems --- from quark-gluon plasma (QGP) created in heavy-ion collisions and primordial matter in the early Universe,  to condensed matter systems such as Weyl and Dirac semimetals. In particular,  the axial anomaly results in axial and electric currents along the background magnetic field --- the chiral magnetic effect (CME) \cite{Vilenkin:1980fu, Fukushima:2008xe, Fukushima:2010vw} and chiral separation effect (CSE) \cite{Son:2004tq, Metlitski:2005pr}.  In the limit of static and uniform fields,  these currents read
\bea
\vec{J}=\frac{\m_5}{2\p^2}\vec{B}~~~,~~~\vec{J}_5=\frac{\m}{2\p^2}\vec{B}\,,
\eea
where $\m$ and $\m_5$ are the vector and axial chemical potentials responsible for the non-zero electric and axial densities.

The structure of the CME indicates that it is non-dissipative \cite{Kharzeev:2011ds,Rajagopal:2015roa, Sadofyev:2015tmb, Stephanov:2015roa} and,  if the electromagnetic (EM) sector is dynamical,  this current may lead to an instability similar to the so-called $\a$-dynamo \cite{Moffatt1978}.  Indeed,  an electric current along the magnetic field results in exponentially-growing helical field configurations supported by the energy stored in the axial charge,  see e.g. \cite{Joyce:1997uy,Akamatsu:2013pjd, Khaidukov:2013sja, Kirilin:2013fqa, Avdoshkin:2014gpa, Manuel:2015zpa, Buividovich:2015jfa, Yamamoto:2015gzz, Hirono:2015rla, Kirilin:2017tdh, Li:2017jwv, Mace:2019cqo, Horvath:2019dvl}. Thus,  such a chiral magnetic instability competes with dissipative processes and may considerably affect the lifetime of magnetic fields generated in non-central heavy-ion collisions or contribute to the dynamics of magnetic fields present in the early Universe.

Another feature of anomalous transport is that chiral media support new collective modes related to these transport phenomena,  see e.g. \cite{Kharzeev:2010gd, Rybalka:2018uzh, Mottola:2019nui, Bu:2018psl}.  In particular,  the interplay between the CME and CSE results in a propagating wave of electric and axial densities,  known as the chiral magnetic wave (CMW) \cite{Kharzeev:2010gd}.  This wave appears as a gapless mode which acquires a mass through the dynamical response of the electromagnetic fields.\footnote{It is argued in \cite{Bu:2018drd,Bu:2019mow} that the CMW may remain gapless due to high-order gradient resummation.}\,The CMW has attracted a lot of attention in the context of heavy-ion collisions since the corresponding charge separation,  while small,  could in principle be detected \cite{Kharzeev:2015znc,Kharzeev:2013ffa,Huang:2015oca}. 

Here we report a novel phenomenon intrinsically related to anomalous transport in chiral matter --- the generation of a chiral dyon by a monopole placed into a chiral plasma.  Magnetic monopoles are often discussed in models of grand unification and evolution of the early Universe,  where they are embedded within the primordial chiral matter.  Effective monopoles may also appear as solitons in QCD and,  moreover,  these objects are actively discussed in models of confinement \cite{Chernodub:2006gu, DAlessandro:2007lae, Shuryak:2018fjr}. For a review of these phenomena as well as the current experimental status of relic monopoles, see e.g.  \cite{Preskill:1984gd, Patrizii:2015uea, Mavromatos:2020gwk}. The generation of a chiral dyon is based on the same interplay between the dynamical EM fields and electric/axial densities in chiral media as in the cases of the chiral magnetic instability and the CMW.  In this setup,  the CMW becomes a spherical wave whose propagation is damped by dissipative currents.  However,  the CME and CSE underlying the CMW in the field of a magnetic monopole lead to a stationary and spherically-symmetric redistribution of charges in the system.  In turn,  the electric charge density affects the electric field through Gauss's law and influences the dissipative currents and anomalous divergence of the axial current,  subject to a detailed balance in the system.  As a result,  the monopole pulls electric charge out of the medium and slightly modifies the nearby axial density,  becoming a chiral dyon. The electric charge of this dyon depends on the averaged axial density and conductivity of the system.

In this paper,  we are guided by a set of simplifying assumptions in order to reveal the principal features of the interplay between a monopole and chiral matter.  We focus on a QED-like theory containing dynamical gauge fields, but non-dynamical axial fields.  The monopole is assumed to have a finite size $a$ with a solid surface at this radius preventing the flow of both in- and out-going currents. We consider uniform hot matter with its temperature $T$ serving as the largest energy scale\footnote{We are working in units where $\hbar = c = k_B = 1$.} in the system, such that $aT\gg 1$, and assume that the electric and axial number densities satisfy $|n|, |n_5|\ll T^3$. This makes it possible to utilize a gradient expansion to lowest order, even in the vicinity of the monopole. While the axial charge is expected to decay due to finite mass effects, we assume that this process is very slow, taking much longer than chiral dyon formation.\footnote{We also note in passing that the CME is argued to be absent in exact equilibrium \cite{Kharzeev:2013ffa, Landsteiner:2012kd, Yamamoto:2015fxa, Zubkov:2016tcp}.}\, The effects due to non-linearities in the densities and/or powers of the magnetic field will be omitted as well. Under these assumptions, linear perturbations in the temperature decouple from the density perturbations and can be ignored. In this simplified model, we find the stationary solutions and show how the axial and electric densities are distributed in space. 

While our work is exploratory and is not aiming at any precision phenomenology,  we provide model estimates of the chiral dyon parameters for monopoles placed into the QGP produced in heavy-ion collisions,  and for the case of the primordial plasma. We also discuss the physical implications of chiral dyon formation, noting that this mechanism may affect the dynamics of the monopoles in the primordial plasma and the process of baryogenesis. Finally,  we note that some condensed matter systems, such as Weyl and Dirac semimetals,  exhibit chiral excitations \cite{Li:2014bha, Huang:2015eia, Xu:2015cga, Lv:2015pya,Li_2016, Randeria:2013kda, 2014ARCMP, Li:2016, Liu:2014} and may be used to detect  relic monopoles through perturbations in 
the electric charge distribution induced by anomalous transport.  

\section{Monopole in Chiral Plasma}

\subsection*{Equations and Solutions}
Our staring point is the linearized constitutive relations for the vector and axial currents, written in the standard hydrodynamic form \cite{landau1959fluid} 
\bea
\label{currents}
&&\vec{j}=-\s\vec{\na}\m+e\s\vec{E}+\frac{e}{2\p^2} \m_5\vec{B}\,,\notag\\
&& \vec{j}_5=-\s\vec{\na}\m_5+\frac{e}{2\p^2} \m\vec{B}\,,
\eea
where $\vec j$ and $\vec{j}_5$ are number currents and $e$ is an elementary electric charge. For simplicity, the diffusion constant in the axial current is assumed to be the same as that in the vector current. The diffusion constant is proportional to the conductivity due to
 Einstein's relation. Since the axial fields are taken to be non-dynamical, the currents are not fully symmetric --- the electric field is generated through Gauss's law by the electric density, while there is no axial electric field.

In the constitutive relations (\ref{currents}), the chemical potentials/densities are assumed to be small and, as such, all non-linear contributions are neglected.  Note that the electric field is of the same order of smallness since it is generated solely by the electric density. While terms simultaneously linear in $\vec{B}$ and in $\m$ or $\m_5$ (or $\vec E$) are maintained, we assume that $\vec B$ satisfies $|\vec B|\ll T^2$ and hence omit terms of cubic order or higher in $\vec B$. Terms quadratic in $\vec B$ unavoidably appear with additional suppression by the electric field or gradients of densities, and they too are omitted. It should be noted that the axial kinetic coefficients, such as the axial conductivity $\s_5$, are forbidden in a P-even theory unless they are proportional to odd powers of $\m_5$ --- such terms are therefore small under our assumptions. Finally, local thermal equilibrium is implied:  linear perturbations in the temperature may enter the currents only through gradients of the temperature multiplied by the corresponding charge density. Such contributions are thus assumed to be negligibly small. 

The dynamics of the system are governed by vector current conservation and the anomalous non-conservation of the axial current: 
\bea
&&\partial_t \,n+\vec \nabla \cdot \vec j=0\,, \notag\\
&&\partial_t \,n_5+\vec \nabla \cdot \vec{j}_5=\frac{e^2}{2\p^2}\vec{E}\cdot \vec{B}\,.
\label{ce}
\eea
While analyzing the coupled dynamics of the system is quite interesting in general, our focus below will be limited to time-independent configurations. In the stationary limit, the continuity equations (\ref{ce}) read
\bea
&&-\s\D \m+e\s\vec{\na}\cdot \vec{E}+\frac{e}{2\p^2}\vec{B}\cdot\vec{\na}\m_5=0\,,\notag\\
&&-\s\D \m_5+\frac{e}{2\p^2}\vec{B}\cdot\vec{\na}\m=\frac{e^2}{2\p^2}\vec{E}\cdot \vec{B}\,,
\eea
where $\D=\vec{\na}\cdot\vec{\na}$ is the Laplacian. These equations are to be completed with the equations for the EM fields and the equation(s) of state relating the densities to chemical potentials. In this spherically-symmetric setup, the magnetic field is defined solely by the monopole charge, while the electric field is determined by the electric density. Correspondingly, Gauss's law reads
\vspace{0cm}
\bea
\vec{\na}\cdot\vec{B}=g\,\delta^{(3)}(\vec{r})~~~,~~~\vec{\na}\cdot\vec{E}=e\,n(\vec{r})\,,
\eea
where one should expect a relation between $g$ and $e$ of the form $e g= 2\p k$ with $k$ being integer (for the Dirac monopole).  Assuming that the medium temperature is the largest energy scale of the system,  we write a simple linearized equation of state
\bea
n=\k\,\m~~~,~~~n_5=\k\,\m_5\,,
\eea
where $\k\sim T^2$ on dimensional grounds and we assume that $|\m|, |\m_5|\ll T$.  The proportionality coefficient depends on the model used for in-plasma interactions,   e.g.  $\k=T^2 /3$  in the case of non-interacting Dirac fermions.

The stationary regime is achieved when the electric current vanishes, as enforced by Maxwell's equations together with the spherical symmetry of the setup $\dot{\vec E}+e\vec J=0$. Thus, in the stationary state there is a detailed balance between the outflow of chirality and its production due to the anomaly. Requiring the electric current to be zero constrains the gradient of the chemical potential. Then, the axial density satisfies
\bea
&&\left[\D-\frac{(a\b)^2}{r^4}\right] n_5(r)=0\,,
\label{n5}
\eea
where $\b$ is a dimensionless combination of parameters controlling the solution, and is given by
\bea
\b=\frac{1}{(2\p)^3}\frac{eg}{a\s}\,.
\eea
This homogeneous equation can be solved analytically, and one finds
\bea
n_5(r)=A_{\rm ch}\cosh\frac{a\b}{r}+A_{\rm sh}\sinh\frac{a\b}{r}\,,
\label{n5sol}
\eea
where $A_{\rm ch}$ and $A_{\rm sh}$ are integration constants to be fixed by the boundary conditions. In turn, the electric density is described by a massive three-dimensional Klein-Gordon equation with an $n_5$-dependent source
\bea
\label{nKGeq}
\left(\D-m^2\right)n(r)=a\b\frac{n_5^\prime(r)}{r^2}\,,
\eea
where $m=e\sqrt{\k}\sim eT$ is the thermal or Debye mass. The solution for this equation can be also written in quadratures
\bea
&&n(r)=\frac{a\,\vare^{mr}}{r} D_++\frac{a\,\vare^{-mr}}{r} D_-\notag\\
&&\hspace{1cm}+\frac{a\b}{m r}\int_a^r\mathrm{d}x\,\sinh{\left[m(r-x)\right]}\frac{n'_5(x)}{x}\,,
\label{eq:nsol}
\eea
where $D_\pm$ are the constants of integration. This general solution can in principle be exponentially-growing at large distances, thus breaking the gradient expansion. However, we will see that the boundary conditions ensure that, in the infinite volume limit, $n(r)$ is in fact a decreasing function whose $r$-gradient is small.  

Since we will be interested in the infinite volume limit, it is instructive to connect (\ref{eq:nsol}) with the Green function of a massive scalar field equation. To do so, one may find it useful to shift the free constants of the homogeneous solution according to
\bea
&&\tilde{D}_+=D_++\frac{\b}{2m}\int^R_a\mathrm{d}x\,\vare^{-m x}\frac{n_5^\prime(x)}{x}\,,\notag\\
&&\tilde{D}_-=D_--\frac{\b}{2m}\int^R_a\mathrm{d}x\,\vare^{-m x}\frac{n_5^\prime(x)}{x}\,,
\label{shift}
\eea
where $R$ is the size of the finite volume spherical system. Then, it is straightforward to show that the solution (\ref{eq:nsol}) takes the form
\bea
\label{nGsol}
&&n(r)=\frac{a\,\vare^{-mr}}{r}\tilde{D}_-+\frac{a\,\vare^{mr}}{r}\tilde{D}_+\notag\\
&&\hspace{1cm}+\frac{a\b}{4\p}\int\displaylimits_{a<|x|<R} \mathrm{d}^3x\frac{\vare^{-m|r-x|}}{|r-x|}\frac{n_5^\prime(x)}{x^2}
\eea
where the Green function can be  easily recognized. From (\ref{nGsol}) it is apparent that the inhomogeneous part of the solution is a falling function of $r$, provided
 the source is not growing too fast at infinity. 

\subsection*{Boundary Conditions}
The free constants are to be fixed by boundary conditions which have yet to be specified. First, as mentioned above, it is important to note that the electric current must be zero in all stationary states of the system. It is sufficient to impose the vanishing of the current at a single radial point, which can be taken to be the monopole radius $a$. We also require no inflow/outflow of axial charge at the monopole surface. This assumption is model-dependent but can be relaxed by taking the point-like-monopole limit. Additionally, the total electric charge $Q$ and axial charge $Q_5$ in the system have to be specified. The corresponding charge densities are used to classify the stationary states of the system. We assume that the plasma is neutral in total, that is  $Q=0$. For any $Q\neq0$, the electric field will push the excess charge to infinity, effectively rendering the system neutral in the bulk. 

Finally, the boundary conditions are given by
\begin{align}
&\int \mathrm{d}^3r\,n_5(r)= Q_5\,,& &n^\prime_5(a)=\frac{\b}{a}n(a),&\notag\\
&\int \mathrm{d}^3r\,n(r)= Q=0\,,& &n^\prime(a)=\frac{\b}{a}n_5(a)\,.&
\label{BCs}
\end{align}

\subsection*{Infinite Volume Limit}
The boundary value problem (\ref{BCs}) is straightforward to solve for arbitrary parameters $\b, m, a$, and $R$. However, it is useful to consider the large volume behavior of the solution, that is  $R\rightarrow\infty$. We start with the equations (\ref{BCs}) for the total charges. The neutrality constraint on the  electric charge sets $\tilde{D}_+=\mathcal{O}\left(\vare^{-mR}\right)$. The remaining terms in (\ref{nGsol}) are then exponentially suppressed,
with (\ref{n5sol}) being inserted  for the axial density. Thus,  returning to (\ref{BCs}), $\tilde{D}_+$ can be set to zero in the rest of the constraints in the infinite volume limit. 

Focusing on the axial charge, there are two possibilities --- either the total charge grows with the size of the system or stays finite. It is convenient to introduce a volume-averaged 
axial charge density $\bar{n}_5$,
\bea
&&\int \mathrm{d}^3r\,\left(A_{\rm ch}\cosh\frac{a \b}{r}+A_{\rm sh}\sinh\frac{a \b}{r}\right)=\bar{n}_5 V_R\,, 
\eea
where $V_R=4\p R^3/3$ is the volume of a sphere of radius $R$. Clearly, if $\bar{n}_5$ vanishes, then $A_{\rm ch}=\mathcal{O}\left(R^{-1}\right)$ and the residual system of constraints for $A_{\rm sh}$ and $\tilde{D}_-$ becomes homogeneous with  a trivial solution only. For $\bar{n}_5\neq0$, implying the scaling $Q_5\sim R^3$, the leading contribution to $A_{\rm ch}$ in the large-$R$ limit  reads
\bea
A_{\rm ch}=\bar{n}_5+\mathcal{O}\left(R^{-1}\right)\,,
\eea
and the system of constraints simplifies considerably. 

The rest of the coefficients can be obtained from the right column of (\ref{BCs}), after $A_{\rm ch}$ and $\tilde{D}_+$ are substituted. The monopole size can be eliminated from consideration by introducing dimensionless units of length, such that $r=a\, \bar{r}$. The stationary states of the infinite chiral plasma in the presence of a monopole are characterized by  two
dimensionless parameters $\b$ and $\g=m a$, as well as the average axial density $\bar{n}_5$. The coefficients read
\begin{widetext}
\begin{align}
\tilde{D}_-&=-\frac{\b \vare^{\g} }{\g}\frac{\g+\b\left(\g\cosh\g-\sinh\g\right)\left(I_{\rm ch}\sinh\b-I_{\rm sh}\cosh\b\right)+\b^2\sinh\g\left(I_{\rm ch}\cosh\b-I_{\rm sh}\sinh\b\right)}{\b^2 \vare^\g I_{\rm ch}+(1+\g)\cosh\b-\b\sinh\b}\bar{n}_5\,,\notag\\
&\hspace{3.6cm}A_{\rm sh}=-\frac{\b^2\vare^{\g}I_{\rm sh}+(1+\g)\sinh\b-\b\cosh\b}{\b^2\vare^{\g}I_{\rm ch}+(1+\g)\cosh\b-\b\sinh\b}\bar{n}_5\,,
\label{eq:coeffs}
\end{align}
\end{widetext}
where we have introduced shorthand notation 
\bea
&&I_{\rm ch}=\int_1^\infty\mathrm{d}x\,\frac{\vare^{-\g x}}{x^3}\cosh\frac{\b}{x}\,, \notag \\
&&I_{\rm sh}=\int_1^\infty\mathrm{d}x\,\frac{\vare^{-\g x}}{x^3}\sinh\frac{\b}{x}\,.
\eea
The coefficients (\ref{eq:coeffs}), together with the constraints on $\tilde{D}_+$ and $A_{ch}$,  provide the full functional dependence of both $n(r)$ and $n_5(r)$ on the parameters of the system. 

\section{Chiral Dyon}

\subsection{Small-$\beta$ Limit}
By looking at \eqref{eq:coeffs}, it is difficult to get an intuitive feel for how the profiles depend on the various parameters, especially due to the presence of the integral functions.  It  is thus instructive to consider the large conductivity limit, which significantly simplifies the expressions. In this limit $\b\ll 1$, the electric field is screened, and the monopole tends to decouple from the plasma. In the next subsection, we will  estimate $\b$ for realistic plasmas and find $\b\ll 1$ to be quite a reasonable approximation for most cases of interest.  Expanding the solutions in powers of $\b$,  the leading contributions to the densities have rather transparent analytic forms
\begin{align}
n(\bar r)/\bar{n}_5&=-\b\frac{\vare^{\g}}{1+\g}\frac{\vare^{-\g \bar r}}{\bar r}+\mathcal{O}\left(\b^3\right)\,,\notag\\
n_5(\bar r)/\bar{n}_5&=1+ \b^2\left( \frac{1}{2\bar{r}^2}-\frac{\g}{1+\g}\frac{1}{\bar{r}} \right) + \mathcal{O}\left(\b^3\right)\,.\label{eq:small-beta}
\end{align}
As expected, the monopole is surrounded by a cloud of electric charge density in a spherical shell with thickness $\d r=1/m$ and with total charge
\bea
\label{eq:ExtraCharge}
q=-\left(\frac{4\p a^3}{\g^2}\b\,\bar{n}_5\right) e=-\left(\frac{1}{2\p^2}\frac{e g}{\s m^2}\bar{n}_5\right) e\,,
\eea
which is non-zero in the limit of a point-like monopole.  In turn, the axial density is also slightly perturbed around the monopole, as can be seen from (\ref{eq:small-beta}), although it cannot be interpreted as being localized in a thin spherical shell.  Thus, a monopole placed into a chiral plasma becomes a chiral dyon. The sign of the dyon's charge is entirely determined by the signs of the axial density $\bar n_5$ and the monopole charge $g$. Overall,  the system maintains electric neutrality as well as its average axial density even in the infinite volume limit,  as can be seen from the solution for any large but finite $R$.  The excess of electric charge,  equal to $-q$,  is pushed to outer spherical boundary,  and the corresponding density tends to zero in the limit of infinite $R$,  while the axial density is both redistributed and partially generated by the anomaly.

Note that the source term in (\ref{nKGeq}) is suppressed with $\b$. Hence the small-$\b$ limit of the solution corresponds to the homogeneous equation,  while dependence on the axial density enters the solution through the near-monopole boundary conditions. These conditions fix the radial currents to vanish at the surface of the monopole, meaning that the diffusion of charges is balanced by the CME and CSE. 

It is instructive to consider the induced electric field and gradients of the obtained densities to confirm that our assumptions are satisfied, at least in this limiting case. Recall that our results are obtained in a linearized approximation, and to the lowest order in gradients. To estimate the latter, we introduce normalized derivatives $f^\prime(r)/(T f(r))$, where $f$ can be either one of the charge densities or the electric field. First note that $r$-derivatives of $n(r)$ always scale either as $1/a$ or $m$. To ensure the validity of the gradient expansion, it is necessary to require that $a\gg 1/T$ and $m\ll T$ --- both constraints are consistent with our approximations.  Next, it is straightforward to see that $n_5^\prime(r)\sim\b^2/a$ --- the derivative is again small for $a\gg 1/T$. Furthermore, it is additionally suppressed by the smallness of $\b$ in this regime.  Finally,  the electric field satisfies $|E(r)/e|<C\left(\b /a^2\right)\left(\bar{n}_5a^3\right)$ with $C\sim 1$ being a numerical factor. Thus, requiring $\bar{\m}_5a$ to not be too large,  where $\bar{\m}_5\sim \bar{n}_5/T^2$, we find the electric field to be small compared to the hydrodynamic scale, $|E(r)/e|\ll T^2$.  In fact, these scalings hold beyond the small-$\b$ limit.

\subsection{Phenomenological Estimates}

\begin{figure*}[t!]
\begin{centering}
\includegraphics[width=8cm]{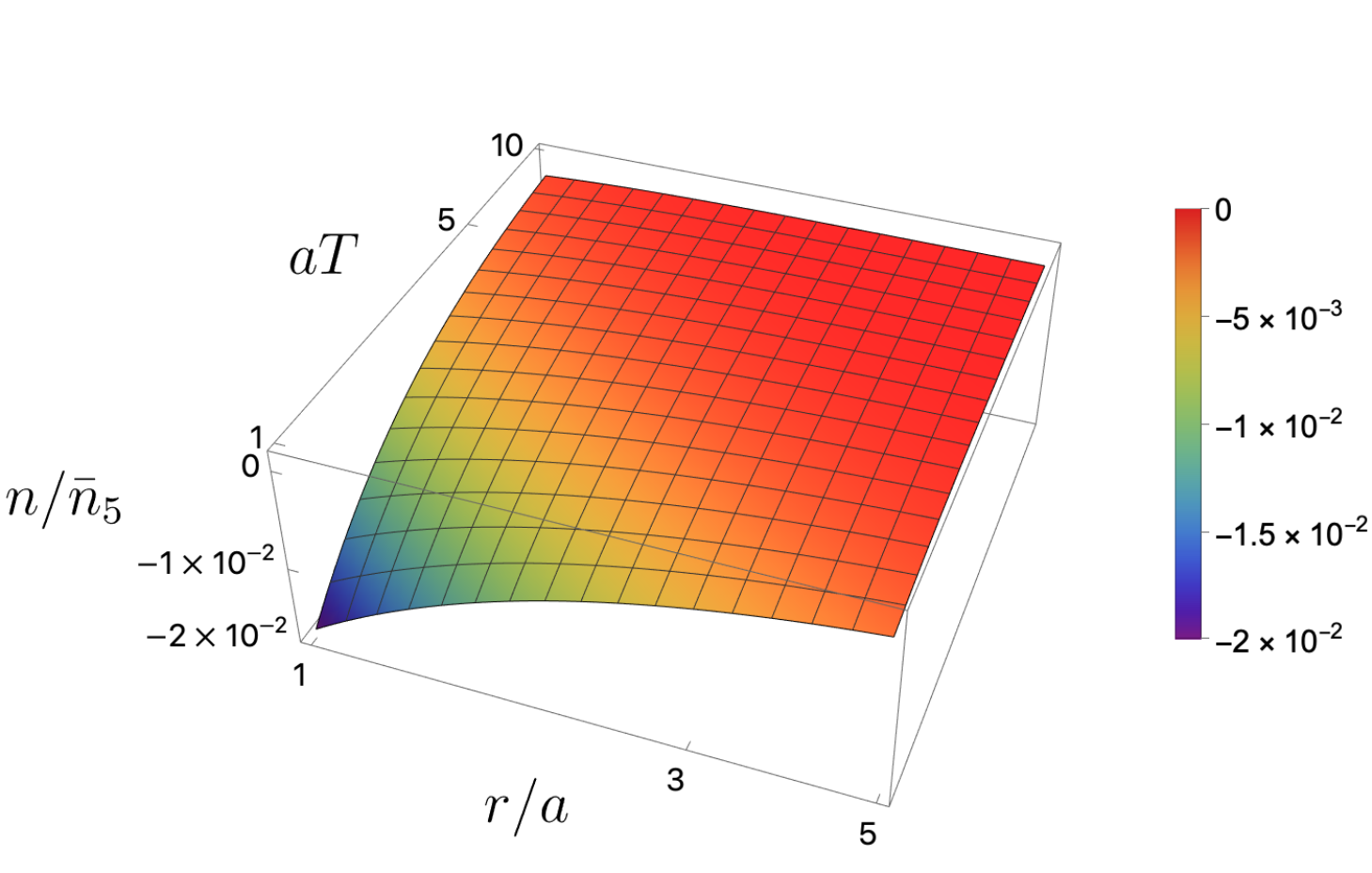}
\hspace{1.5cm}
\includegraphics[width=8cm]{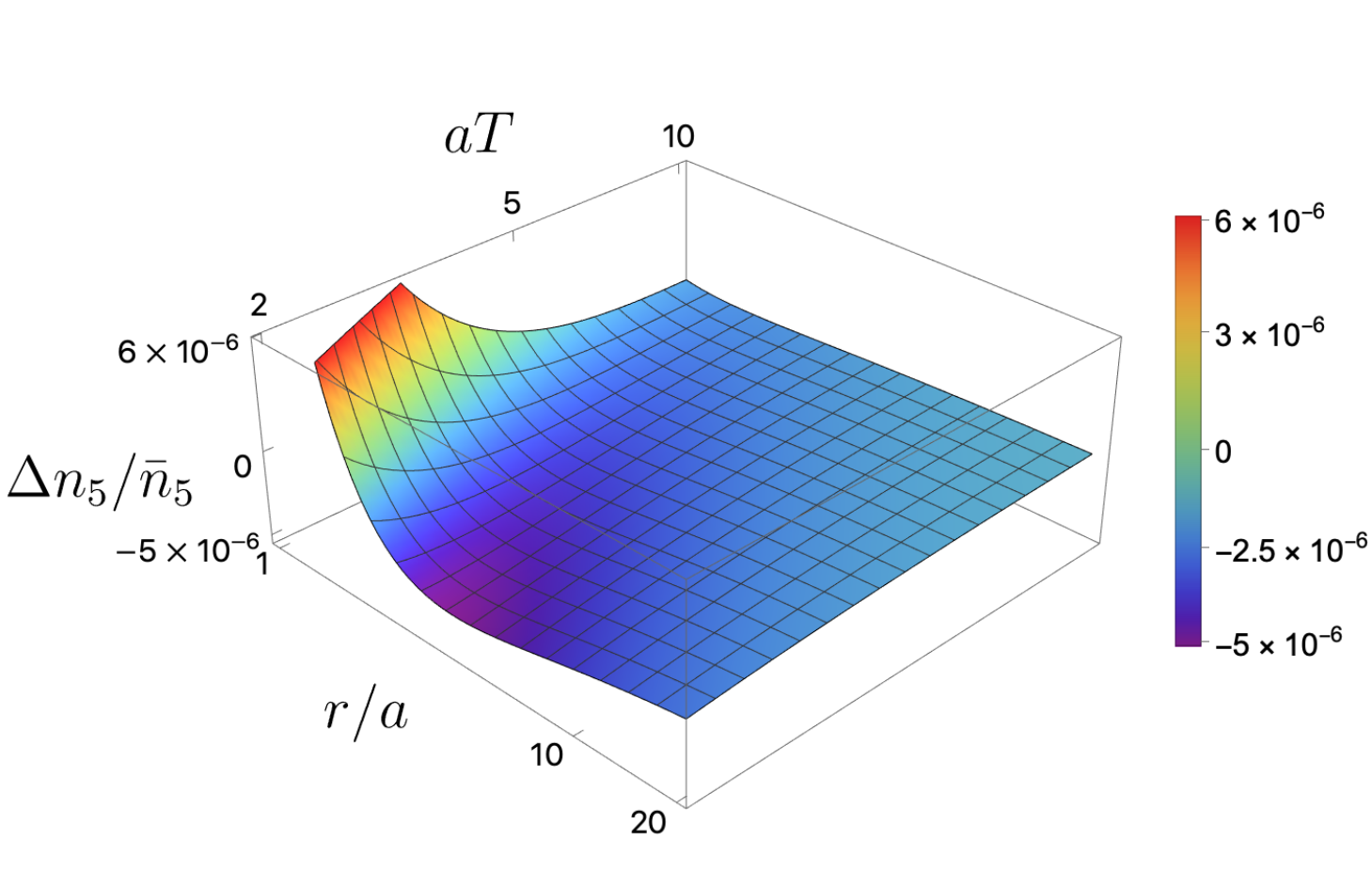}
\end{centering}
 \caption{(Left) The electric density as a function of normalized temperature and radial distance.  (Right) The offset axial density as function of normalized temperature and radial distance.  In these plots we set $\bar{\s}=10^{-1}$,  a change in its value results in an overall rescaling of the plotted function. The offset axial density is cut off at a value of $6\times 10^{-6}$ for the purposes of presentation.}
 \label{fig:nANDn5-T}
\end{figure*}

Let us now roughly estimate the magnitude of the effect that the presence of a monopole has on the charge density in some realistic examples of chiral media. Eq.~\eqref{eq:coeffs} reveals that the precise dependence of $n$ and $n_5$ on the free parameters $\g$ and $\b$ can be quite complicated. The parameters themselves are temperature-dependent and, in fact, the temperature is the dominant parameter controlling the setting and final results. Let us first consider the case of free Dirac fermions at high temperature by setting\footnote{Obviously,  considering $eg=2\p k$ with larger $k$ will scale the result accordingly.} $eg=2\p$ and substituting $\bar n_5=T^2\bar\m_5/3$ into (\ref{eq:ExtraCharge}). This gives
\bea
\frac{q}{e}=-\frac{1}{\p\bar\s}\frac{\bar{\m}_5}{T}\,,
\eea
where we have introduced the dimensionless variable $\bar\s=e^2\s /T$.  For the nearly-chiral plasma in the early Universe (at $T\sim 10-100\,\text{GeV}$), the characteristic conductivity\footnote{The conductivity $\s$ enters the number current and is normalized such that a single power of $e$ enters in front of it.  To compare to most of the literature,  $\s$ has to be rescaled by $e^2$.} can be estimated to be $\bar\s\sim 140$ in our model containing a single fermion flavor \cite{Baym:1997gq,Ahonen:1998iz,Arnold:2000dr}.  In contrast, the temperatures of QGP produced in heavy-ion experiments are much lower, roughly below
$1\,\text{GeV}$.  The characteristic value of $\bar\s$ in QGP is around $\bar\s\sim 10^{-2}-10^{-1}$ at $T=T_c\sim 160\, \text{MeV}$ \cite{Puglisi:2014sha}. While the axial chemical potential is expected to be smaller than the dominant energetic scale given by the temperature, $q$ is maximized for $\bar{\m}_5\sim T$.  Thus $q/e<1$ in the early Universe, that is, the electric charge of the chiral dyon is smaller than a single unit of the elementary electric charge.  This is true unless $\bar{\m}_5$ is considerably larger than the temperature.  In contrast,  the charge of the dyon in the QGP phase may reach $q\sim e$  if the axial chemical potential remains an order of magnitude smaller than the temperature, leaving  room for even higher charges.

It is worth noting that the plasma under discussion is not necessarily weakly interacting.  For instance,  in the  limit of a strongly-coupled holographic plasma, the only major modification to the discussion above is a change in the kinetic coefficients in (\ref{currents}).  If the elementary ``electric'' charge $e$ is normalized to $e=1$, then $\k= 2(\pi T)^2 $ and $\bar\s= \pi$ in the large-$T$ limit, see e.g. \cite{Policastro:2002se,CaronHuot:2006te,Erdmenger:2008rm,Rajagopal:2015roa,Bu:2015ame}.  The only free parameter in such a holographic model of a strongly interacting plasma is the anomalous coefficient, which enters through the chiral effects and the anomaly itself.

From (\ref{eq:ExtraCharge}) one can see that the total charge surrounding the monopole is independent of $a$,  while the density profile is  sensitive to its value.  Considering the two dimensionless parameters, we find that they scale as
\bea
&&\b=\frac{1}{(2\p)^3}\frac{eg}{a\s}\simeq\frac{0.002}{\bar\s \: (a T)}\notag\\
&&\g=ma\simeq0.17(a T)\,,
\eea
where we again use $\k=T^2/3$. For the purposes of estimation,  the classical monopole size may be taken to be equal to the classical electron radius. Following this choice and setting $a\sim 1\,\text{fm}$,  $\b$ appears to be small for temperatures larger than $0.002/(a\bar{\s})\simeq 0.4\,\text{MeV}/\bar{\s}$. Thus in most cases, the small-$\b$ expansion may be safely used to describe the system, and one can therefore rely on (\ref{eq:coeffs}), (\ref{eq:small-beta}), and (\ref{eq:ExtraCharge}) for such purposes.\footnote{The notion of a monopole size is delicate.  For lower temperatures,  quantum effects caused by large EM fields appear inside the classical electron radius.  At higher temperatures, these effects are in fact suppressed compared to in-matter thermal fluctuations,  up to distances of order $1/T$.  Thus one may make naive estimates using $a \sim 1/T$,  which is, however, beyond the accuracy of the present work.  We expect the observed picture to largely hold even inside the monopole radius. For more accurate estimates one must consider the monopole field beyond its classical structure, as used throughout this work, and take non-linear effects into account.}

\begin{figure*}[t]
\begin{centering}
\includegraphics[width=10cm]{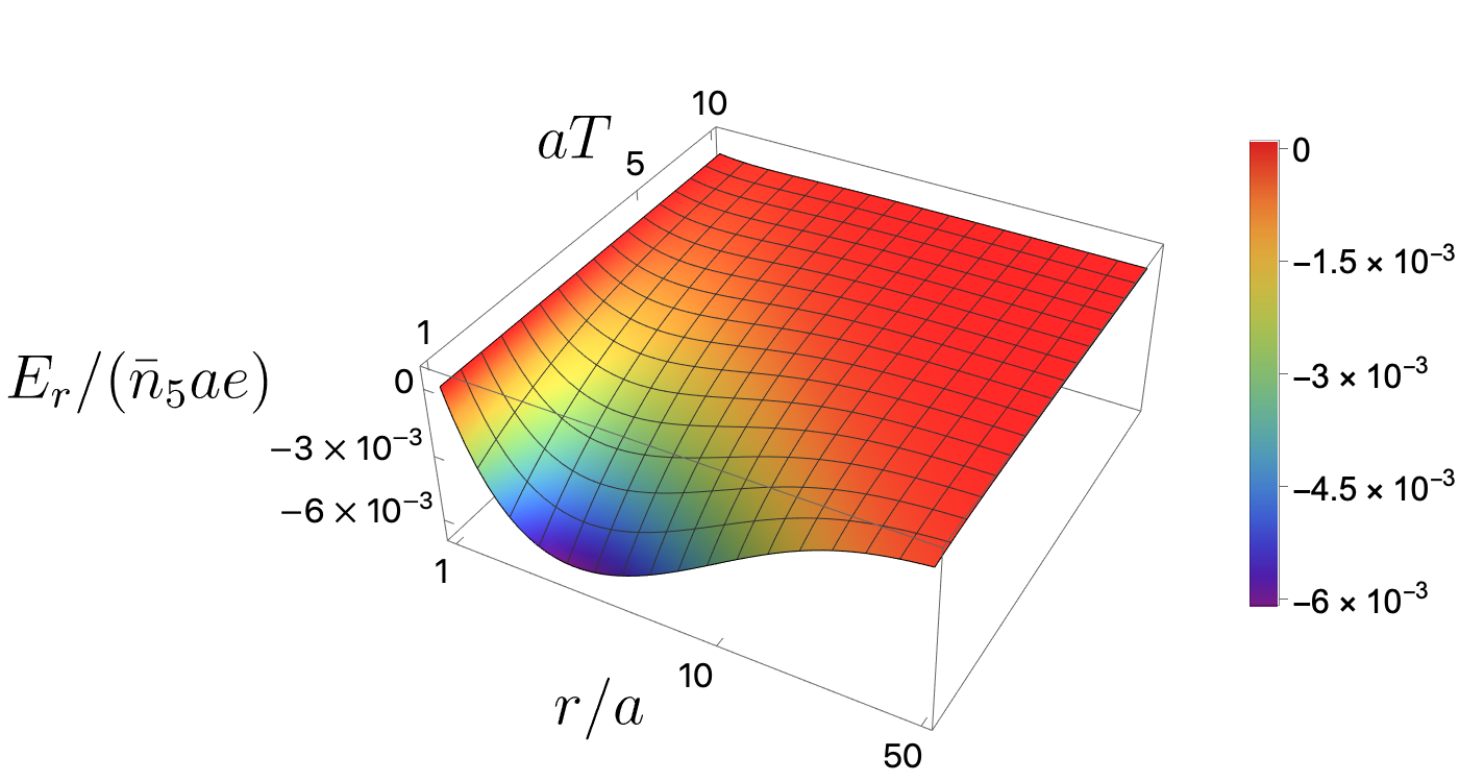}
\end{centering}
 \caption{The electric field $E_r$ as a function of normalized temperature and radial distance. As in the previous figure, we set $\bar{\s}=10^{-1}$.}
 \label{fig:E-T}
\end{figure*}

The estimates for $\g$ and $\b$ are now employed in order to plot the radial profiles for the normalized electric density $n(\bar{r})/\bar{n}_5$,  the offset axial density $\D n_5/\bar{n}_5 \equiv n_5(\bar r)/\bar{n}_5 -1$,  and the electric field $E_r$, all as functions of the temperature $T$ and radial distance $r$.  The axes are normalized by the monopole radius $a$. The temperature is allowed to go as low as $a T=1$, slightly stretching the assumptions made in our consideration.  For these plots we set $\bar{\s}=10^{-1}$. Since $\b=0.02/(aT)\ll1$,  the full solutions are  well-described by (\ref{eq:small-beta}). Fig.~\ref{fig:nANDn5-T} (left) represents the electric density, whose shape is dominated by the decaying exponent. Fig.~\ref{fig:nANDn5-T} (right) represents the offset axial density, where the shapes of projections to fixed values of $(aT)$ depend on the relative sizes of the two $\mathcal{O}\left( \b^2 \right)$ terms in (\ref{eq:small-beta}).  For $\g>1$,  which corresponds roughly to the region of large $(a T)$,  $\D n_5/\bar{n}_5$ is negative for any $\bar r$,  while for $\g<1$, the offset density changes sign at $\bar r\simeq(1+\g^{-1})/2$.  Note that the offset axial density is plotted up to an upper cutoff set at $6\times 10^{-6}$ in order to provide better resolution of the sign-changing region.  Finally,  the electric field profile due to the generated electric density is presented in Fig.~\ref{fig:E-T}.

\section{Discussion}
In this paper we have reported on the effect produced by a single magnetic monopole when it is inserted into a chiral medium at finite temperature.  The radial magnetic field of the monopole electrically 
polarizes the medium, forming a chiral dyon with an electric charge that is dependent on the chiral asymmetry of the matter.  This phenomenon is governed by an interplay between the CME,  CSE,  and EM field dynamics, and in this sense it is similar to the CMW and chiral magnetic instability. 

Our present work is exploratory and can be expanded in several directions,  all left for future investigations.  
First,  since only the stationary solutions are found and presented here,  one might question the stability of the solutions
with respect to time-dependent perturbations, or search for new dynamical solutions to (\ref{ce}). 
Next, it would be desirable to relax some of the approximations made.  In particular, if the monopole radius $a$ 
is taken to be small,  then the gradient expansion would break in the vicinity of the monopole, and one would have to employ gradient resummation in the spirit of \cite{Bu:2015ame, Bu:2016oba, Bu:2016vum}.  
Furthermore,  the EM fields near the monopole would become strong and some non-linear effects might need to be included \cite{Bu:2018psl, Bu:2018drd}.  Finally, realistic chiral plasmas are frequently out of equilibrium, and one might not be able to rely on the near-equilibrium hydrodynamic description.  The formalism of chiral kinetic theory \cite{Arnold:2002zm, Stephanov:2012ki, Son:2012zy, Gorbar:2016qfh} could be instrumental in the study of chiral matter polarization by a monopole,  beyond the hydrodynamic regime. The interplay between a probe monopole and chiral matter within the chiral kinetic theory approach has been considered in \cite{Yamamoto:2020phl},  though that work does not consider the generation of the charge asymmetries or the formation of the chiral dyon.

Above, we discuss the case of a single monopole with exact spherical symmetry.  However,  one may consider
a more physical monopole-antimonopole configuration.  Intuitively,  both will turn into chiral dyons with opposite electric charges,  resulting in mutual attraction.  Such a system would be electrically unstable and eventually collapse in the absence of other forces.  A more realistic scenario would be a monopole-antimonopole gas.  For such a gas,  one could use the single-monopole solution discussed above with a finite system-size $R\sim n_M^{-1/3}$,  with $n_M$ being the monopole density \cite{Lublinsky:2009iu}. 

Moreover,  one might consider the case of a purely-monopole gas (or monopole-antimonopole plasma with an excess of monopoles).  Phenomenologically,  this is quite an interesting case since, due the mechanism reported in this work,  the monopole excess should result in the generation of electric charge. In a more realistic theory --- a theory in which the particle content of the chiral medium is specified --- this can be translated into either baryonic or leptonic charge asymmetry.
We are lead to speculate that through the formation of chiral dyons, magnetic monopoles in the early Universe  can,  in principle,  contribute to the process of baryogenesis.
Furthermore, the dyons disturb the distribution of charge and axial densities, and can affect time evolution of the monopole density itself.

A natural generalization of the effect discussed above would be that of a dyon placed into a chiral plasma.  In fact,  under the same assumptions, the sole result would be the modification of the boundary conditions at the monopole surface, due to  the additional contribution to the electric current.  Indeed,  since the electric field cancels through the derivation of (\ref{n5}),  the equation for $n_5$ is left unmodified,  while the electric density equation (\ref{nKGeq}) gains no new contributions for $r>a$.  Since the boundary problem is linear, the dyon can be considered as a combination of two solutions: one for a monopole and the other for an electric charge.  As an electric charge placed into conducting matter gets fully screened,  the case of the dyon differs from the case of the monopole only by a slight redistribution of the densities.  The full electric charge of the dressed dyon is equal to that given by (\ref{eq:ExtraCharge}) in the limit of large conductivity. However, one should be careful in regards to potential non-linear effects since, due to the Dirac quantization condition, either the electric or magnetic field of the dyon can be large. 

Finally,  we would like to note that there is a completely different class of systems supporting chiral fermions --- Weyl and Dirac semimetals.  In these systems,  the characteristic parameters are expected to differ considerably from those previously discussed and,  moreover, to some extent be controllable by a smart choice of the material. Provided that a regime of sufficiently large chiral imbalance is achievable,  these new materials could be used to detect relic monopoles through the formation of chiral dyons.  Such dyons would have the potential to be experimentally observed as slowly propagating clouds of charge.  For instance,  one may expect that  formation of the chiral dyon will result in a higher energy loss of the monopole in chiral media.  Furthermore,  over the last years there has been quite a lot of research activity aiming at using topological materials (including Dirac and Weyl semimetals) as dark matter detectors, see e.g. \cite{Hochberg:2017wce,Marsh:2018dlj}. There might be an opportunity to tune such experiments to simultaneously search for magnetic monopoles.

\section*{Acknowledgements}
The authors would like to thank A. Badamshina,  who participated at early stages of this study.  The authors are grateful to D. S. M.  Alves and M.  Graesser for helpful discussions. The work of ML is supported by the Israeli Science Foundation (ISF) Grant No.~1635/16 and the BSF Grant  No.~2018722.  The work of JR is supported by the UC Office of the President through the UC Laboratory Fees Research Program under Grant No.~LGF-19-601097.  The work of AS on Sec.  I and Sec.  II is supported by the Russian Science Foundation Grant RSF 21-12-00237 and on Sec.  III and Sec.  IV by the European Research Council project ERC-2018-ADG-835105.  AS is also grateful for support from Xunta de Galicia (Centro singular de investigaci\'on de Galicia accreditation 2019-2022),  from the European Union ERDF,  from the Spanish Research State Agency by “Mar{\'{i}}a de Maeztu” Units of Excellence program MDM-2016-0692,  project FPA2017-83814-P,  and from European Union’s Horizon 2020 research and innovation program under the Grant Agreement No.~82409.

\bibliographystyle{bibstyle}
\bibliography{monopole}

\end{document}